\documentstyle[preprint,revtex]{aps}
\begin{document}


\begin{title}
Anisotropic Pairing on a
Three-Band Fermi Surface for $A_3{\rm C}_{60}$
\end{title}

\author{E. J. Mele and S. C. Erwin}
\begin{instit} Department of Physics and
Laboratory for Research on the  Structure of Matter\\University of
Pennsylvania, Philadelphia,  Pennsylvania 19104 \end{instit}

\begin{abstract}
	We study a model for anisotropic singlet pairing in $A_3{\rm C}_{60}$,
using a realistic model for the Fermi surface in a hypothesized
orientationally-ordered doped crystal. Anisotropic solutions are studied by
combining numerical solutions to the gap equation in the low-temperature
phase with a Landau expansion near the mean field critical temperature.  We
focus on a class of three-dimensional nodeless $d$-wave solutions to the
model,  which exhibit a fully developed gap everywhere on the Fermi
surface, but non-BCS temperature dependence of the order parameter in the
condensed state, a relatively large value of the low-temperature gap,
$2\Delta/kT_{\rm c}$ , and non-BCS structure in the quasiparticle spectrum near
the
gap.

\end{abstract}
\pacs{PACS numbers: }
The discovery of superconductivity in
the alkali-doped fullerenes \cite{ross91} has raised a number of
interesting questions about the interplay of  superconductivity and the
unique structural properties exhibited by these systems. Exchange of intra-
and intermolecular phonons of the crystal have both been examined as
possible pairing mechanisms for this system \cite{schl92,varm91,zhan91a}.
These
studies have considered a simple pairing scenario in which the condensation
is presumed to occur on a Fermi surface consisting of a single sheet, with
the resulting pair amplitude taken to be symmetric under all the point
group symmetry operations of the crystal.  This choice is conceptually
simple and furthermore it is apparently motivated by a growing body of
experimental work on the condensed state which indicates that the gap is
nonvanishing everywhere on the Fermi surface \cite{uemu91} .

In this paper we develop a theory of pairing in $A_3{\rm C}_{60}$ which
treats the structure of the Fermi surface of this system more
realistically.  The model we develop below  is actually relevant to an
orientationally-ordered doped system, and in this sense remains highly
idealized.  Nevertheless, we find that even at this level the structure and
topology of the three-band Fermi surface  of  this system has important
effects on the pairing dynamics.  In particular, we discuss a class of
stable, anisotropic, but nodeless solutions to the gap equation for the
three band model, in which  an ``axial'' paired state is constructed from a
complex combination of the anisotropic pairing fields of $e_g$ symmetry.
Interestingly, this solution does not exhibit either the line or point
nodes expected of an anisotropic state owing to a very interesting
interference between this pair field and critical points which occur at
intersections between the various sheets of the Fermi surface. In addition
to breaking the $T_h$ symmetry of the high-temperature phase of the
doped crystal, this phase is distinguished by a non-BCS temperature
dependence of the order parameter in the condensed state, a relatively
large value of the low temperature gap,  $2\Delta/kT_{\rm c}$ , and non-BCS
structure in the quasiparticle spectrum near the gap.  It is an interesting
speculation that, unlike the mean field critical temperatures that are
predicted for the related fully symmetric $a_g$ ($s$-wave) paired state,
these results will prove to be relatively resistant to the effects of the
short range interelectronic repulsive interactions that are omitted from
our treatment. However, the fate of this phase in the presence of disorder
remains a very open and relevant question.

Here we  briefly present our model Hamiltonian, discuss the symmetry of the
low temperature pair field, discuss our numerical low temperature solutions
to the gap equation, and develop a Landau theory for the phase transition
near the critical temperature.  A number of details of this treatment will
be developed in an expanded discussion to be presented elsewhere.

The conduction band of the alkali-doped fullerenes is derived from an
orbitally three fold degenerate $t_{1u}$ state of the isolated molecule.
The hopping amplitude between nearest neighbor sites can thus be
represented by a 3$\times$3 matrix, and to describe it we adopt the
parameterization of Gelfand and Lu \cite{gelf92} , so that
$H_t=t\Psi^{\dag}_{i\mu\sigma} [ T_{\mu\nu}(i,j) ] \Psi_{j\nu\sigma}$
where $\Psi^{\dag}_{i\mu\sigma}$ creates an electron with spin $\sigma$ in
orbital  $\mu$  on site $i$. The conduction-band width is then controlled by
the scale parameter $t$,  while the symmetry of the conduction states is fixed
by the form of the nearest neighbor 3$\times$3 real matrices $T(i,j)$ given in
\cite{gelf92}.   Energies in this paper will all be measured in units of $t$;
the full bandwidth of the conduction band in the orientationally ordered doped
state is $W=42 t$.

The Fermi surface (FS) of the model at half filling (three electrons per
primitive cell) is shown in Fig.\ 1 \cite{erwi91}. The surface consists of
two pieces.  One surface forms an empty hole pocket at the Brillouin-zone
center.  The second surface is constructed from pairs of parallel tubes
which run along the Cartesian $\langle 001 \rangle$ directions.  Symmetry
requires that these two sheets of  FS ``touch'' at single points along the
$\langle 111 \rangle$ directions approximately given by 0.52 $(\pi /a)
\langle 111 \rangle$; this  symmetry will play a central role in the
discussion that follows.

We now expand this model to include an interaction term $H_{int}$ which is
decoupled by a BCS factorization which yields:
\begin{equation} H_{int}=\sum_{n{\bf k}\mu\nu\sigma} \Psi^{\dag}_{{\bf
k}\mu\sigma}
D_{n,\mu\nu} \Psi^{\dag}_{-{\bf k}\nu-\sigma} \langle \Delta_{n{\bf k}} \rangle
+
\langle \Delta^{\dag}_{n{\bf k}} \rangle  \Psi_{-{\bf k}\mu\sigma} D_{n,\mu\nu}
\Psi_{{\bf k}\nu-\sigma} - \langle \Delta^{\dag}_{n{\bf k}} \rangle
\frac{1}{U_n}  \langle \Delta_{n{\bf k}} \rangle  \end{equation} with $\langle
\Delta_{n{\bf k}} \rangle = U_n \langle \Psi_{-{\bf k}\mu\sigma}
D_{n,\mu\nu}\Psi_{{\bf k}\nu-\sigma}\rangle$. Here the
$\Psi^{\dag}_{{\bf k}\mu\sigma}$ creates a Bloch orbital with momentum {\bf k}
and
the 3$\times$3
matrices $D$    specify the the orbital symmetry of the pair fields that are
coupled through the  interaction potentials $U_n$; a condensation in the $n$th
channel may occur when $U_n < 0$.     For the fully symmetric state with
$D_1={\bf {\rm I}}$, exchange of phonons of $A_g$ and $H_g$  symmetry
both contribute to an attractive potential $U_1$.   The electrons can also
be paired in an $e_g$ doublet which is spanned by two  basis functions with
$D_{2,3}  =
\frac{1}{\sqrt{6}}\left(\begin{array}{rrr}1&0&0\\0&1&0\\0&0&-2\end{array}\right),
\frac{1}{\sqrt{2}}\left(\begin{array}{rrr}1&0&0\\0&-1&0\\0&0&0\end{array}\right)$
.  Basis functions for pairing  in a $t_{2g}$ orbital triplet can also be
constructed, but a calculation of pairing susceptibility in this channel shows
that it is much smaller at half filling and we will not consider it further.
We
will consider only the effect of a static (non-retarded) attractive potential,
with the one-electron bandwidth then providing a high energy cutoff on the
theory.

We begin by studying the second-order pairing susceptibilities
$\partial^2\Omega/\partial\Delta_\alpha \partial\Delta^{\ast}_{\alpha} = \{
1/| U| - \chi_\alpha(T)\}$, calculated in this model as a function of
temperature $T$ for the various pair fields listed above when the
conduction band is half filled.  The $\chi$'s  are all logarithmically
divergent at low temperature. In the on-site approximation, $U_1\equiv U_2
\equiv U_3$, we find that at a given temperature
$\chi_\alpha$  is largest for the fully symmetric $a_g$ pair field. In the
$e_g$ channel $\chi$ is only slightly smaller so that the form of the
paired state for this system will depend sensitively on the relative sizes
of the of the interaction potentials $U_n$.   For a simple model which
includes only intra-ball scattering and ignores the direct repulsion
between electrons, these are predicted to be exactly equal for the symmetric
and anisotropic channels, so that condensation in the symmetric channel
should be weakly favored in such a theory.  Nonetheless, the anisotropic
solution is a locally stable solution to the model, and it seems to provide
a better representation of some aspects of the experimental data.
Moreover, it can be dynamically stabilized in any theory for which $(3/2)
| U_1| < | U_2 |$, and so we now examine its properties in more
detail.

The zero temperature ${\bf k}$-integrated quasiparticle spectra near the gap
are
shown in Fig.\ 2.  The top panel is obtained for the fully symmetric
paired state. and shows a characteristic BCS singularity in the spectral
density of the form   $1/\sqrt{(E-\mu)^2-\Delta^2}$. The light curve in the
second panel is obtained for a real anisotropic pairing field of $e_g$
symmetry.  Here, the quasiparticle spectrum exhibits a pseudogap with the
spectral density vanishing as a power law at the Fermi energy and  peaking
at  logarithmic critical points located deeper in the band; these can
already be resolved in these numerical data.   However, the ``real''
gapless state considered in Fig.\ 2(b) is not the lowest-energy paired
configuration in the $e_g$ manifold;  a lower free energy is obtained
through the axial combination $D_\pm = (1/\sqrt{2})(D_2\pm iD_3)$.  Below we
show that this combination minimizes  the fourth-order terms in a Landau
expansion of the free energy near the transition temperature, so that the
phase transition from the normal state occurs directly to the axial phase.
This is also established by a numerical solution to the gap equations which
we have carried out in the low temperature phase, and which indicate that
the axial phase has a lower free energy than the simple  ``real'' pairing
states of $e_g$ symmetry.  The reason for this is apparent in Fig.\ 2(b)
which shows that the axial combination (heavy curve) stabilizes a ground
state with a well-developed non-vanishing gap everywhere on the Fermi
surface.

Nodeless $d$-wave pairing in a ``two-dimensional'' system is also possible, and
has been discussed in the context of high $T_{\rm c}$ superconductivity in the
cuprates.  For this situation the full three-dimensional Fermi surface is
presumed to take  the topology of a cylinder that opens along the $k_z$
axis, so that the complex combination $\Delta({\bf k}) =(k_x \pm ik_y)^2$
yields a
non-vanishing pair amplitude of $d$-wave symmetry which then neatly avoids the
point zeroes at the poles.  The three-dimensional $A_3{\rm C}_{60}$ system
can not make use of this escape route, and employs a rather more devious
strategy to avoid the point zeroes that would otherwise be expected for
the axial phase.

In Fig.\ 1 we have also displayed the trajectories of the nodes of the
order parameter, calculated as the zeroes of the diagonal matrix elements
$\langle \Psi_{n{\bf k}} D_\alpha \Psi_{n-{\bf k}} \rangle$ for the
$\alpha$th pair state on
the $n$th band crossing the Fermi level. On each surface these nodal
trajectories are on a collision course and are required to meet at each of the
high symmetry points where the FS crosses the $\langle 111 \rangle$ directions.

Along the $\langle 111 \rangle$ directions in ${\bf k}$-space, however, the
situation is  more interesting.  Here the one-particle kinetic energy
operator $H_t$ takes the  form   $H_t = \left( \begin{array}{ccc}
e({\bf k})&v({\bf k})&v({\bf k})\\v({\bf k})&e({\bf k})&v({\bf k})\\ v({\bf
k})&v({\bf k})&e({\bf k})\end{array}\right)$,
 so that the $t_{1u}$ molecular state is broken into a  two-fold degenerate
$\Lambda_3$ state  and a nondegenerate $\Lambda_1$
 state.  The crossing of this $\Lambda_3$ doublet through the Fermi
surface  marks the  ``contact point'' between the inner and outer sheets of
the Fermi surface.  The pairing amplitude at this critical point can  be
written as a real traceless  2$\times$2 matrix for each of the $e_g$
components.
These matrices are not independent since  in the $\Lambda_3$  subspace
$\langle D_2\rangle\equiv (1/2)({\bf {\rm I}}+i\sigma_y)\langle
D_3\rangle({\bf {\rm I}}-i\sigma_y)$.   Nonetheless, the eigenvalues of $D$ can
only vanish if
both the diagonal and off-diagonal components go to zero simultaneously
exactly at the crossing point. This cannot occur since $[H_t,D]$ is
non-vanishing along the $\Lambda$ line,  so that $H_t$ and $D$ cannot
simultaneously be rotated to diagonal form.  Physically, the off-diagonal
components of the pairing tensor dominate at the crossing point and provide
the required non-vanishing pair amplitude.

The order parameter of this axial phase exhibits a non-BCS temperature
dependence below $T_{\rm c}$.  This is shown in Fig.\ 3(a)  where the
temperature dependence of the pair amplitude $\langle\Delta_\pm\rangle$,
and $E_{\rm c}$, the energy of the logarithmic critical point (which would be
readily identified in any ${\bf k}$-integrated measurement) are plotted as
functions of temperature. For comparision, the standard weak-coupling BCS
prediction is shown as the dashed curve.  If we identify the logarithmic
singularity in these spectra as the ``gap'' for this system, a relevant
parameter which can  then be extracted from this plot is
$E_{\rm c}/kT_{\rm c}=2.3$,
which is considerably larger than the familiar weak coupling BCS prediction
of 1.76.  In fact particle and hole amplitudes along the crucial $\Lambda$ line
are extremely strongly mixed even out to an energy $E_\Lambda/kT_{\rm c}$ =
3.08,
which we believe sets the true energy scale for the pairing in this
state.    Nevertheless thermal excitations should exhibit an activated
temperature dependence  governed by the {\it minimum} gap, which is seen to
be substantially smaller for this phase.

Also shown in Fig.\ 3(a) is the temperature dependence of
$\langle\Delta_2\rangle$  and $\langle\Delta_3\rangle$, which have the same
mean field critical temperature, but are stabilized by larger fourth-order
coefficients in the Landau expansion, and develop weaker ground-state
expectation values at low temperature.  For comparison in Fig.\ 3(b) we
also display the results of  calculations for pairing in the fully
symmetric channel (this is obtained on the same temperature scale by
reducing the coupling strength $U$ by 2/3 in the symmetric channel) .
Although the transition temperatures are comparable,
$\langle\Delta_1\rangle$  is found to follow the expected weak-coupling BCS
temperature dependence very  closely.

The transition from the paramagnetic phase to the axial phase can be
studied by developing a Landau expansion in the order parameters
$\Delta_2$ and
$\Delta_3$.  To quadratic order the free energy of this system takes the form:
$\Omega=\Omega_0 + (1/| U| - \chi) (|\Delta_2|^2 +
|\Delta_3|^2) = \Omega_0 + a[(T-T_{\rm c})/T_{\rm c}](|\Delta_2|^2 +
|\Delta_3|^2)$.   The symmetry of the condensed state is then
determined by the fourth-order invariants in the Landau expansion, which
can  be grouped as $\Omega_4=g_1[|\Delta_2|^2+|\Delta_3|^2]^2 +
2g_2|\Delta_2|^2|\Delta_3|^2 + g_3(\Delta_2^2\Delta_3^{\ast 2}
+  \Delta_2^{\ast 2} \Delta_3^2)$.  We have evaluated each of the
coefficients $g$ by a perturbation expansion from the high-temperature
paramagnetic phase carried to fourth order.  Stability requires that $g_1>
0$ and cubic symmetry requires that $g_2 + g_3 - g_1 = 0$.  If we
parameterize $\Delta_2=\Delta_3 e^{i\phi}$,  we see that this potential
then takes the simple form
$\Omega_4=g_1[|\Delta_2|^2+|\Delta_3|^2+
2g_3|\Delta_2|^2|\Delta_3|^2(\cos 2\phi - 1)$, so that the
symmetry of the condensed state is specified solely by the sign of $g_3$.
For positive $g_3$, we see that $\Omega$  is minimized with
$|\Delta_2|=|\Delta_3|$ and $\phi=\pm\pi/2$,   which is exactly
the axial phase.  Conversely, for negative $g_3$ the system orders with
$\phi = 0$, and in this case the potential is invariant under an arbitrary
continuous rotation in the $\Delta_2-\Delta_3$ plane. (This continuous
symmetry is not a symmetry of the cubic crystal and is ultimately broken at
sixth order in the Landau expansion.)  Numerically we find that $a =
0.017/t$, $g_1 = 0.0086/t^3$ ,  and  $g_3 = 0.0029/t^3$. The positive $g_3$
indicates that the axial phase is indeed favored over the ``gapless''
$e_g$ phases for this system and confirms the results of the low
temperature analysis given above.

It is difficult to make a detailed comparison betwen this simple model and
experimental data.  Nevertheless, we do note that the temperature
dependence of the gap has been studied by Lieber and coworkers using
scanning tunneling spectroscopy \cite{zhan91b}, and this yields a classical
mean-field ``weak coupling'' temperature dependence of the gap in the ordered
state, but with a non-BCS ratio $\Delta/kT_{\rm c}=2.6$. This result has been
interpreted as evidence of extremely strong coupling in this system, a
result which is surprising (as the authors of \cite{zhan91b} note)  since
this gap is known to be a relatively small fraction of the phonon
frequencies generally  thought to be relevant for scattering the
conduction electrons at the FS.  Infrared reflectance measurements also
yield a relatively large gap, $\Delta/kT_{\rm c}=1.5-2.5$, and this has been
interpreted as resulting from a distribution of gaps in experimentally
inhomogeneous disordered materials \cite{rott92}.  The temperature
dependence of the muon-spin relaxation rate has cleanly ruled out an order
parameter with nodes on  the Fermi surface \cite{uemu91} which, as we have
seen, is consistent with pairing in either the fully symmetric or axial
channels.

The most provocative property of the axial phase is that it exhibits chiral-
symmetry breaking in the pair field.   This symmetry breaking, if found in
this system,  should manifest itself in any macroscopic property that
derives from the low-lying quasiparticle excitations.  We should point out
that for the simple pairing model studied here, the global minimum of the
free energy is actually found for a condensation in the fully symmetric
$a_g$ channel.  The influence of more realistic potentials on this result
and of the  effect of orientational disorder, which seems to be an
unavoidable experimental difficulty for the doped fullerenes at present,
remain to be explored.

This work was supported by the DOE under Grant 91ER45118, and by the
NSF-MRL under Grant 91 20668.
%
%

%
%
\figure{The two sheets of the Fermi Surface in the three-band model of
$A_3{\rm C}_{60}$. The outer surface is cut away to reveal  the inner
surface.  The lines denote zeroes of the pairing fields for the $\Delta_2$
and $\Delta_3$  states (light and heavy curves, respectively). \label{fig1}}

\figure{Quasiparticle spectral densities near the gap for pairing in the
fully symmetric $a_g$ state (a), for a real $e_g$ state (b, light curve) and
for the axial  $e_g$ state (b, heavy curve). \label{fig2}}

\figure{Temperature dependence of the order parameter, and prominent
features of the spectral densities for pairing states of different
symmetries. In (a) the filled circles denote $\langle\Delta_\pm\rangle$,
and the filled squares
denote $E_{\rm c}$, the location of the logarithmic peak in the density of
states
for this phase. For comparison,we include the expectation values in the
real states $\langle\Delta_2\rangle$ (diamonds) and
$\langle\Delta_3\rangle$ (open circles).  The dashed line is
the weak-coupling BCS result normalized to the mean-field transition
temperature.  The chain-dot curves are the results of Landau expansions in
the BCS model and in our anisotropic model near $T_{\rm c}$.  In (b) the filled
circles denote $\langle\Delta_1\rangle$ for the fully symmetric channel,
and the filled squares
are the locations of the peaks in the spectral density, which follow the
weak-coupling BCS model (dashed curve) quite closely. \label{fig3}}

\end{document}